\documentclass[aps,prb,twocolumn,showpacs,10pt,superscriptaddress]{revtex4-1}
\usepackage{graphicx,amsmath,amssymb,amsfonts,sidecap,color}

\usepackage{dsfont}
\usepackage{graphicx}
\usepackage{subfigure}
\usepackage{amsmath}
\usepackage{amsfonts}
\usepackage{amssymb}
\usepackage{mathrsfs}
\usepackage{bbm}
\usepackage{subfigure}
\usepackage{xcolor}
\usepackage[colorlinks=true]{hyperref}

\begin{document}

\title{Landau-Zener-St\"{u}ckelberg Interferometry for Majorana Qubit}

\author{Zhi Wang}
\affiliation{School of Physics, Sun Yat-sen University, Guangzhou 510275, China}
\affiliation{International Center for Materials Nanoarchitectonics (WPI-MANA)\\
National Institute for Materials Science, Tsukuba 305-0044, Japan}

\author{Wen-Chao Huang}
\affiliation{School of Physics, Sun Yat-sen University, Guangzhou 510275, China}

\author{Qi-Feng Liang}
\affiliation{Department of Physics, Shaoxing University, Shaoxing 312000, China}
\affiliation{International Center for Materials Nanoarchitectonics (WPI-MANA)\\
National Institute for Materials Science, Tsukuba 305-0044, Japan}

\author{Xiao Hu}
\email{Hu.Xiao@nims.go.jp}
\affiliation{International Center for Materials Nanoarchitectonics (WPI-MANA)\\
National Institute for Materials Science, Tsukuba 305-0044, Japan}

\begin{abstract}
Stimulated by a very recent experiment observing successfully two superconducting states
with even- and odd-number of electrons in a nanowire topological superconductor
as expected from the existence of two end Majorana quasiparticles (MQs)
[Albrecht \textit{et al.}, Nature \textbf{531}, 206 (2016)], we propose a way to manipulate Majorana qubit exploiting quantum tunneling effects. The prototype setup consists of
two one-dimensional (1D) topological superconductors coupled by a tunneling junction which can be controlled by gate voltage.
We show that, upon current injection, the time evolution of superconducting phase difference at the junction induces an oscillation in energy levels of the Majorana parity states,
whereas the level-crossing is avoided by a small coupling energy of MQs in the individual 1D superconductors. This results in a Landau-Zener-St\"{u}ckelberg (LZS) interference
between the Majorana parity states. Adjusting the current pulse and gate voltage, one can build a
LZS interferometry which provides an arbitrary manipulation of the Majorana qubit. The LZS
rotation of Majorana qubit can be monitored by the microwave radiated from the junction.
\end{abstract}
\date{\today}
\maketitle

%74.50.+r (Tunneling phenomena; Josephson effects),
%03.65.-w (Quantum mechainics),
%03.67.-a (Quantum information),
%74.90.+n (Other topics in superconductivity),
%74.81.Fa (Josephson junction arrays and wire networks),
%76.63.Nm (Quantum wires),
%74.20.Mn (Nonconventional mechanisms),
%85.25.Cp (Josephson devices), 85.25.Dq (SQUIDs)}

\section{Introduction}
Zero-energy quasiparticle excitations in topological superconductors behave like Majorana fermions \cite{Majorana,ReadGreen,Kitaev,ivanov}.
Significant efforts have been made to realize Majorana
quasiparticles (MQs) in recent years\cite{kanermp,aliceareview,beenakkerreview,kanefu,law09,sarma1d,oppen1d,aliceanphy,kouwenhoven,pergescience,Sau11,LiangWangHu12,WuLiangHu14}, with the expectation that these illusive quasiparticles obeying non-Abelian statistics can be used to build
decoherence-free quantum computation\cite{Kitaev,ivanov,kanermp,aliceareview,beenakkerreview}. Although MQs are charge neutral as they are constituted of
electron and hole with equal weights (thus the equivalence between particle and antiparticle), two of them can compose a complex fermion, which specifies
superconducting states carrying either odd- or even-number of electrons.
The two parity states can be exploited to build a qubit, which is stable against local electromagnetic noises. Upon braiding the MQs
the system can be driven from an eigenstate with definite parity
into a superposed state with quantum entanglement\cite{Kitaev,ivanov,kanermp,aliceareview,beenakkerreview,aliceanphy,kanefu,Sau11,LiangWangHu12,WuLiangHu14}.
However, it is well known that the weights and phases of superposed Majorana qubit states cannot be controlled arbitrarily by braiding operations only.
For quantum computation, a universal gate enabling a complete control on MF qubits is required, which breaks the topological protection with minimal invasive perturbation\cite{bonderson10,hassler10,jiang,flengsberg11,pekker13,schmidt}.

In a very recent experiment\cite{albrecht}, degenerate superconducting states with even- and odd-number of electrons in a nanowire topological superconductor of mesoscopic size have
been revealed in terms of the Coulomb blockade effect\cite{Fu,WangHuLiangHu}.
Exponentially small energy differences between the two states are observed originated from a small overlapping between wavefunctions of MQs at the
two ends of a nanowire superconductor when the parallel magnetic field is tuned away from the matching condition among the chemical potential, external magnetic field
and the superconducting gap function for realizing the zero-energy MQs\cite{sarma1d}.
This experimental breakthrough motivates us to seek for the possibility of building a compact universal gate for Majorana qubit in the nanowire system.

\begin{figure}[t]
\begin{center}
\includegraphics[clip = true, width = 1\columnwidth]{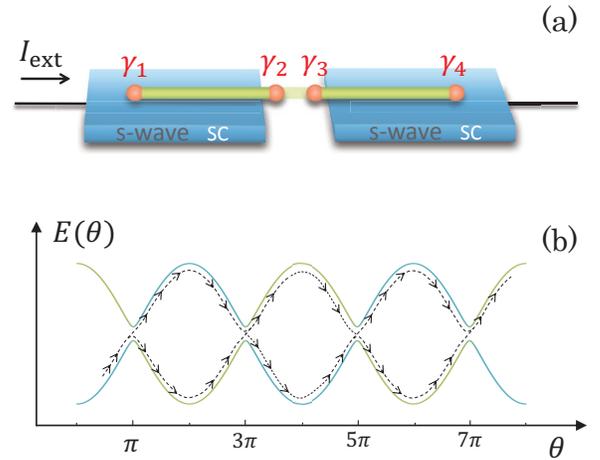}
\caption{(Color online) (a) Schematic of a quantum tunneling junction between two 1D topological superconductors driven by a bias current $I_{\rm ext}$.
(b) Energy levels of the two eigenstates as function of phase difference $E(\theta) = \pm \sqrt{E^2_m\cos^2(\theta/2) + \delta^2}$
(see text), with the blue and green curves indicating the two pure parity states. The system in a parity eigenstate is scattered into a superposition
of two parity states upon passing through $\theta=\pi$, and after many periods of oscillation in the superconducting phase difference evolves into the parity state opposite to the initial one
as the manifestation of LZS interference.}
\end{center}
\end{figure}

We study a prototype setup of Majorana qubit involving a quantum tunneling junction between two one-dimensional (1D) topological superconductors as sketched schematically in Fig.~1(a) achieved by laying nanowires of semiconductor with large spin-orbit coupling on the $s$-wave superconductors and applying a parallel magnetic field (not shown explicitly)\cite{sarma1d,albrecht,aliceanphy}.
Injection of current into the system induces a voltage drop across the junction, which makes the superconducting phase difference between the two superconductors
evolve with time according to the ac Josephson effect\cite{Josephson1,Josephson2},
and meanwhile generates small but appreciable interactions between MQs in individual segments
similarly to the case when the parallel magnetic field is tuned in experiments\cite{albrecht}.
Solving a Shr\"{o}dinger equation for the even- and odd-parity states of the Majorana qubit, we find intriguingly
that the Majorana qubit rotates with time as a manifestation of Landau-Zener-St\"{u}ckelberg (LZS) interference\cite{landau,zener,stueckelberg}, with a frequency proportional to the small MQ interactions in individual 1D topological superconductors. Furthermore
we demonstrate that one can control accurately the weights of the even- and odd-parity states by adjusting the pulse length of bias current, which, along with the control on the MQ coupling at the junction by gate voltage,
achieves an arbitrary manipulation of the Majorana qubit. The scheme is scalable at least linearly. The LZS
rotation of Majorana qubit can be monitored by the microwave radiated from the junction.

\section{LZS interference of Majorana parity states}
As revealed by previous works\cite{sarma1d,schmidt}, the two MQs residing at the junction couple to each other in the form of $iE_{\rm m}\gamma_2\gamma_3$ through
single-electron tunneling, where the energy $E_{\rm m}$ can be controlled by a gate voltage.
Because the total parity is conserved, the quantum state of the system sketched in Fig.~1(a) is described by the following Hamiltonian\cite{leggett,WangLZS},
\begin{eqnarray}
H_{\rm m} = E_{\rm m} \cos(\theta/2) \sigma_z + \delta \sigma_x,
\label{hamiltonian}
\end{eqnarray}
with basis of the even- and odd-parity state $i\gamma_2\gamma_3|0,1\rangle=\pm |0,1\rangle$,
where $\sigma_{x,z}$ are Pauli matrices and $\theta=\phi_1-\phi_2$ is the phase difference
between the two superconductors. The interaction term $\delta$ between the two Majorana parity states is due to the couplings between MQs in individual 1D topological
superconductors [between $\gamma_1$ and $\gamma_2$ in the left one and between $\gamma_3$ and $\gamma_4$ in the right one in Fig.~1(a)],
which, while being negligibly small at the pristine condition, is enhanced to an appreciable (but small) value when a voltage drop is induced at the junction by current injection. 
For the purpose of manipulating the Majorana qubit, the junction is tuned by the gate voltage such that
$\delta\ll E_{\rm m}<\Delta$, where $\Delta$ is the gap function of proximity-induced superconductivity in the nanowires. In this situation, the eigenstates of the system
are the superpositions of the two Majorana parity states around $\theta=(2n+1)\pi$ with the relative weights
determined by the MQ interaction $\delta$, whereas coincide with the two pure parity states elsewhere [see Eq.~(\ref{hamiltonian}) and Fig.~1(b)].

When the current passing through the junction is beyond the critical current of the junction,
the superconducting phase difference starts to evolve with time, which triggers a quantum mechanical evolution of the Majorana qubit as can be read from Eq.~(\ref{hamiltonian}).
This dynamics can be described by the following the time-dependent Schr\"{o}ding equation
\begin{eqnarray}
i \hbar \frac{d}{dt} \left[\begin{array}{cc}
\psi_0  \\
\psi_1
\end{array}\right] = \left[\begin{array}{cc}
E_{\rm m}\cos{\frac{\theta(t)}{2}} & \delta\\
\delta & -E_{\rm m}\cos{\frac{\theta(t)}{2}}
\end{array}\right] \left[\begin{array}{cc}
\psi_0  \\
\psi_1
\end{array}\right],
\label{schrodinger}
\end{eqnarray}
with $\Psi=\psi_0|0\rangle + \psi_1|1\rangle$.

\begin{figure}[t]
\begin{center}
\includegraphics[clip = true, width = 0.9 \columnwidth]{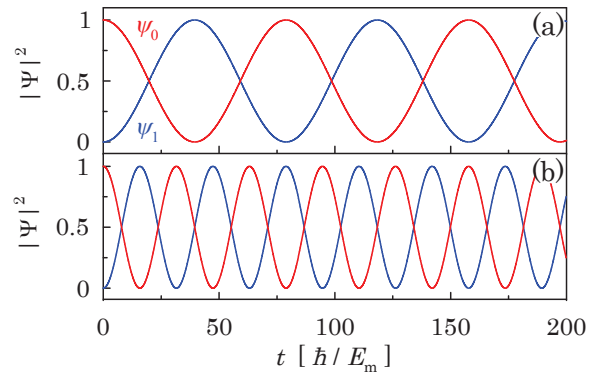}
\caption{  LZS interference between the two parity states with (a) $\delta/E_{\rm m} = 0.04$ and (b) $\delta/E_{\rm m} = 0.1$.
The frequency of oscillation in the superconducting phase difference induced by the bias current is $\omega=30E_{\rm m}/\hbar$.
}
\end{center}
\end{figure}

When the system is driven dynamically through $\theta=(2n+1)\pi$, the Majorana qubit may either stay in its original parity state, or evolve
into the opposite parity state [see Fig.~1(b)]. The probability for staying in the same state is given by the
ratio between the crossing-avoid energy and the velocity of the phase variation $\propto \exp [-\delta^2/(\hbar\omega E_{\rm m})]$\cite{landau,zener}.
Experiencing many passages through the crossing points, the occupation probabilities of the two parity states are governed by the
accumulation of quantum phases acquired at individual passages known as the LZS interference\cite{landau,zener,stueckelberg,nori},
which yields a second, longer time scale as compared with the period of the oscillation of superconducting phase difference.

The dynamical process can be revealed by integrating numerically Eq.~(\ref{schrodinger}). As shown in Fig.~2, upon a fast oscillation of
phase difference induced by current injection $\theta(t)=\omega t$,
the Majorana qubit rotates between the even- and odd-parity states, with the rotating frequency proportional to the value of $\delta$, manifesting the LZS interference.

An analytic description of the quantum mechanical evolution of the Majorana qubit is available resorting to the Floquet theorem, which provides a general theoretical scheme
for treating any system with time-periodic hamiltonian\cite{floquet,shirley}. For $\delta=0$
the time-dependent Schr\"{o}dinger equation (\ref{schrodinger}) can be transformed to the time-independent one with the Floquet matrix hamiltonian of infinite dimensions corresponding to
the photon numbers\cite{son}. The effect of the small MQ interaction $\delta\ll E_m, \omega$ can then be taken into account perturbatively.
To the first-order perturbation approximation which involves a $2\times 2$ Floquet matrix, we obtain a LZS oscillation in occupation probabilities
$|\psi_0(t)|^2=\cos^2(\omega_{\rm m}t/2)$ and $|\psi_1(t)|^2=\sin^2(\omega_{\rm m}t/2)$ with
\begin{equation}
\omega_{\rm m}= \delta J_0(4E_{\rm m}/\hbar\omega)/\hbar
\label{LZSfrequency}
\end{equation}
starting from the initial state $\psi_0(0)=1$ and $\psi_1(0) = 0$, where $J_0(x)$ is the Bessel function.

The frequency of Majorana qubit rotation is linearly proportional to the interaction between the two parity states as can be understood intuitively.
In contrary, the frequency of oscillation in superconducting phase difference $\omega$ is involved in a nonlinear way, and does not influence much the Majorana qubit
rotation in the large limit because the reduced probability for parity flipping at a single passage is compensated by the increasing number of passages.
For the case of Fig.~2(a), Eq.~(\ref{LZSfrequency}) gives $\omega_{\rm m} \approx 0.0398 E_{\rm m}/\hbar$ taking into account
 $\omega = 30E_{\rm m}/\hbar$, and numerical simulations give $\omega_{\rm m} = 0.0399 E_{\rm m}/\hbar$, whereas for the case of Fig.~2(b), one has
$\omega_{\rm m}\approx 0.0996 E_{\rm m}/\hbar$ analytically and $\omega_{\rm m} = 0.09955 E_{\rm m}/\hbar$ numerically, agreeing well with each other in both cases.

\begin{figure}[t]
\begin{center}
\includegraphics[clip = true, width = 1 \columnwidth]{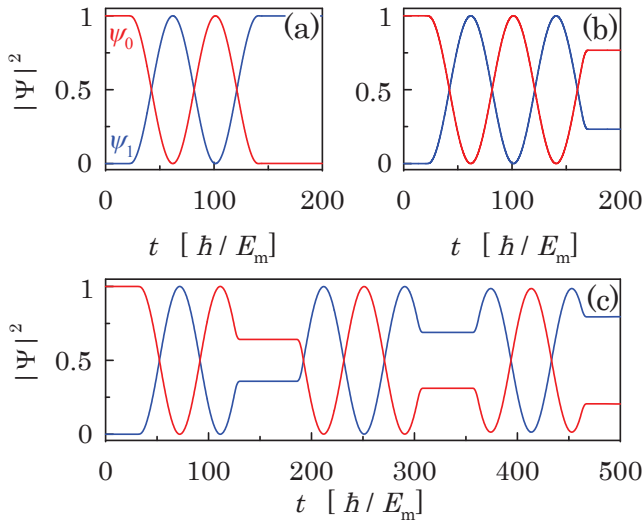}
\caption{LZS interferometry on the Majorana qubit by bias-current pulse, with a short pulse (a) and a long pulse (b), and a sequence of three
pulses (c). Parameters are taken the same as Fig.~2(a) except for $\omega=0$ and $\delta=0$ when the bias current is turned off. }
\end{center}
\end{figure}

\begin{figure}[t]
\begin{center}
\includegraphics[clip = true, width = 1 \columnwidth]{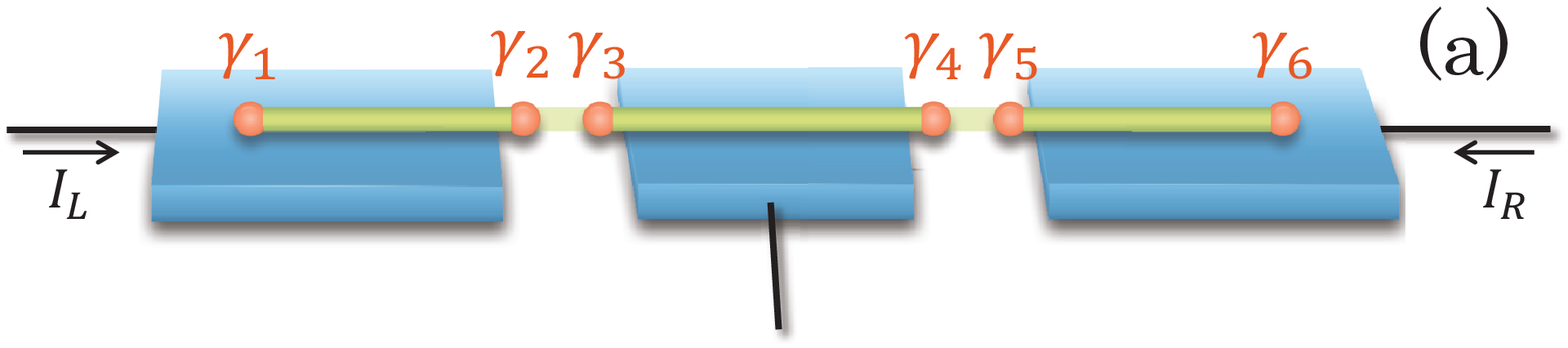}
\includegraphics[clip = true, width = 1 \columnwidth]{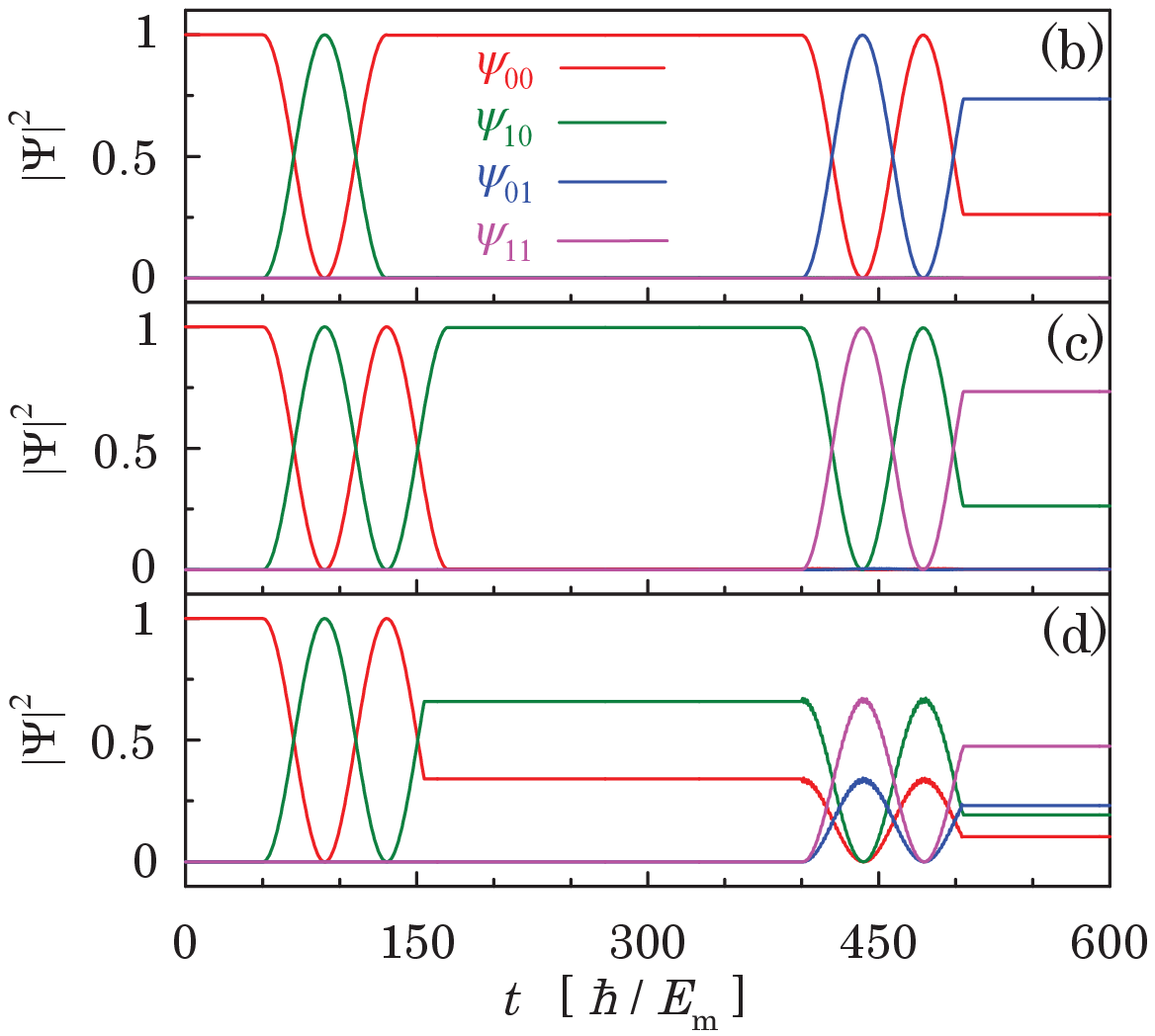}
\caption{ (a) Schematic for LZS interferometry for two Majorana qubits. (b)-(d) a sequence of controls on
the left Majorana qubit with the right Majorana qubit staying static, and vice versa.  In the first operation, the bias current in the left/right junction is switched on/off, and
$E_{\rm m, L}= E_{\rm m}$ and $E_{\rm m,R}=2 E_{\rm m}$ are taken; then bias current is switched off in both junctions
for a time period during which the energies $E_{\rm m,L}$ and $E_{\rm m, R}$ are switched; in the second operation the bias current in the left/right junction is switched off/on.
Parameters are taken the same as Fig.~3.}
\end{center}
\end{figure}

\section{LZS interferometry for Majorana qubit}
Knowing that the Majorana qubit can rotate completely between the even- and odd-parity states, it is intriguing to check whether one can control the MF qubit arbitrarily.
For this purpose, we investigate the response of the system to pulses of bias current.

As shown in Figs.~3(a) and (b), we observe that the occupation probabilities in the two Majorana qubit states start to oscillate when the current is turned on, and the
oscillations stop when the bias current is turned off. Rotation between the two Majorana qubit states can either continue or switch back after turning on again the current bias, as seen in Fig.~3(c) at the second and third current pulse respectively.
Therefore, weights of the even- and odd-parity states can be tuned arbitrarily by adjusting the length of
current pulse.

Since scalability is important for practical implementation of qubit gate, we investigate a system with two qubits as schematically shown in
Fig.~4(a), where an electrode is attached to the superconductor at the middle such that current can be injected only into one of the two junctions at a time.
The time-dependent Schr\"{o}ding equation for the system of two Majorana qubits takes the same form as Eq.~(\ref{schrodinger}) except for that the Hamiltonian
matrix is $4\times 4$ with the diagonal entries $\epsilon_j=\pm E_{\rm m,L}\cos(\theta_{\rm L}/2)\pm E_{\rm m,R}\cos(\theta_{\rm R}/2)$ standing for the energies of the four parity states with
$j=|00\rangle, |10\rangle, |01\rangle, |11\rangle$, where $E_{\rm m,L}$ and $E_{\rm m,R}$ are the two Majorana coupling constants at the left and right junctions respectively, and
off-diagonal entries $\delta_{ij}$ standing for the couplings among the qubit states.
Because the two Majorana $\gamma_3$ and $\gamma_4$ are residing on the same 1D topological
superconductor [see Fig.~4(a)] and thus are intrinsically coupled to each other, at the first glance it would seem impossible to control the two Majorana qubits separately
even though no current passes directly through one of the two junctions.  A hint for overcoming this difficulty can be found in Fig.~1(b) that a Majorana qubit changes
its parity state only when the two states are close to each other in energy in the order of MQ interaction $\delta$.
We therefore assign the two coupling energies at the junctions to satisfy $|E_{\rm m,L}-E_{\rm m,R}|\gg \delta$.
As shown in Figs.~4(b) and (c), in the first operation the left Majorana qubit rotates with the right one remaining static, and verse versa in the second operation; one can
stop the left Majorana qubit at either the even-parity state [Fig.~4(b)] or the odd-parity state [Fig.~4(c)]. One can also choose to stop the left Majorana
qubit at a superposed state of the even- and odd-parity states as shown in Fig.~4(d).
This observation indicates unambiguously that the weights of parity states at the two Majorana qubits can be controlled independently.

In order to manipulate the phases of the four Majorana qubit states, we can turn off
the bias current and control the timing for switching off the junction coupling $E_{\rm m}$,
which rotates the phases of the Majorana qubit states to desired values [see Eq.~(\ref{schrodinger})].
Experimentally this can be achieved by controlling the gate voltage at junction as addressed by the previous work\cite{schmidt}.
Therefore, an arbitrary control on Majorana qubits can be achieved in terms of adjusting current injections and gate voltages at junctions.

\begin{figure}[t]
\begin{center}
\includegraphics[clip = true, width = 0.9 \columnwidth]{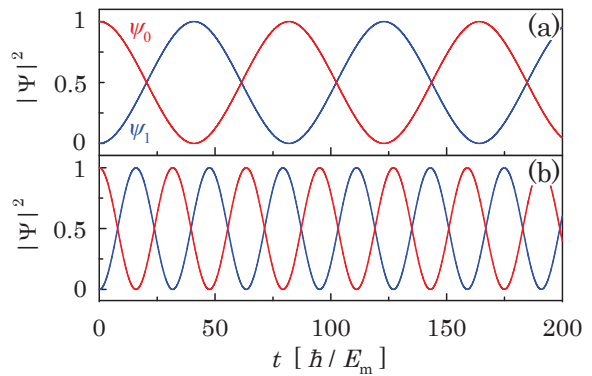}
\caption{(Color online). LZS interference between the two parity states upon current bias obtained by solving simultaneously Eqs.~(\ref{schrodinger}) and (\ref{rcsj}) with
$R = 8{\hbar}/{2e^2}$, $C = 0.04{e^3}/{\hbar I_{\rm s}}$, $I_{\rm m} = 0.5I_{\rm s}$ for which $I_{\rm c} \approx 1.375 I_{\rm s}$, $I_{\rm ext} = 3I_{\rm c}$,
and (a) $\delta/E_{\rm m} = 0.04$ and (b) $\delta/E_{\rm m} = 0.1$.}
\end{center}
\end{figure}

\section{RCSJ phase dynamics}
LZS interferometry has so far been proposed for manipulating superconducting flux qubits and Cooper-pair box qubits\cite{son,nori,GGC}.
The present proposal has a potential advantage that only current injection is required which can be performed quickly as compared with an operation involving magnetic flux.
In order to demonstrate that the LZS interferometry for Majorana qubit discussed above can be achieved stably, we estimate
the magnitude of bias current. For this purpose, we analyze the
resistively-and-capacitively-shunted dynamics of superconducting phase difference at the junction\cite{tinkham,husust}
\begin{equation}
I_{\rm ext}=\frac{\hbar C \ddot \theta}{2e} + \frac{\hbar \dot \theta}{2eR} +I_{\rm s}\sin{\theta} +I_{\rm m} (|\psi_0|^2 - |\psi_1|^2) \sin\frac{\theta}{2},
\label{rcsj}
\end{equation}
where $C$ and $R$ are the effective capacitance and resistance of the junction [see Fig.~1(a)], and $I_{\rm s}$ and $I_{\rm m}$ are the threshold currents for the channels of
Cooper pairs and MQs respetively\cite{Kitaev,aliceanphy}. This semi-classical treatment on superconducting phase difference is possible since
it evolves with time much faster than the Majorana qubit [see Eq.~(\ref{LZSfrequency})]. For the present purpose, the
current injected through the junction should be larger than the critical value
$I_{\rm c} = (2I_{\rm s} \zeta  + I_{\rm m}) \sqrt{1- \zeta^2}$, with $\zeta =\sqrt{({I_{\rm m}}/{4I_{\rm s}})^2 + 1/2} - {I_{\rm m}}/{4I_{\rm s}} $, which induces a
voltage drop at the junction and drives the time evolution of superconducting phase difference according to the ac Josephson effect\cite{Josephson1, Josephson2}.
Conventionally the RCSJ dynamics of the superconducting phase mimics a Newtonian particle moving in a tilted washboard potential\cite{tinkham,husust}.
In the present case, the phase particle acquires an additional pseudospin degree of freedom associated with the Majorana parity states as described by Eq.~(\ref{rcsj}).
The LZS oscillation of the Majorana qubit changes the supercurrent through the $4\pi$-period channel, and then recoils to influence the time evolution of superconducting phase difference.
Therefore, the full dynamics of the system should be derived by solving simultaneously Eqs.~(\ref{schrodinger}) and (\ref{rcsj}).
As shown in Fig.~5, for a bias current $I_{\rm ext}= 3 I_{\rm c}$ corresponding to $\omega\simeq 30E_{\rm m}/\hbar$,
the time evolutions of Majorana qubit derived with the RCSJ dynamics
and those obtained by presuming a linear phase dynamics shown in Fig.~2 agree well with each other. It is because that under this bias current the
superconducting phase difference evolves with time sufficiently fast such that the phase particle hardly senses the washboard potential. Therefore, one can conclude that the LZS interferometry for the Majorana qubit can be implemented in terms of moderate bias current through a junction with small critical supercurrent.

The oscillation of superconducting phase difference at the junction upon current injection radiates a microwave\cite{husust}. As can be read from
Eq.~(\ref{rcsj}), there is a new spectral weight induced by the MQ tunneling around half value of the frequency determined by the ac Josephson effect,
which exhibits a peak splitting due to the rotation of Majorana qubit. The noise spectrum of the microwave radiation can therefore be
used for monitoring the quantum mechanical dynamics of the MQs.
While the resistance does not appear explicitly in the results discussed so far, it would induce decoherence of the quantum state since it passes normal, dissipative current.
Nevertheless, for the value of resistance given in Fig.~5, no appreciable decoherence can be observed over one hundred rotations of Majorana qubit,
meaning that decoherence caused by the normal component of current is sufficiently weak.  The operation time without destroying the quantum coherence of Majorana qubit
can be prolonged by choosing a larger resistance at the junction.

\section{Discussions}
In a previous work\cite{schmidt}, a microwave cavity was introduced to control the MQ coupling at a junction between two 1D topological superconductors, which in addition with a braiding operation completes the universal gate of Majorana qubit. In the present work, we show that, making use of the LZS interferometry, this goal can be achieved simply by the gate control at the junction and current injection along the nanowires.

Finally we give typical numbers for relevant physical quantities. The topological superconducting gap is in the order of 200~$\mu$eV according to the state-of-art
experiments\cite{kouwenhoven,albrecht}, whereas the
coupling of MQs at the junction should be smaller by one order or so, say $E_{\rm m} \sim 20~\mu$eV. The coupling between MQs in individual 1D topological superconductors, the smallest energy
scale of the present system, is $\delta\sim 5~\mu$eV typically. In this case the
LZS frequency reads $\omega_{\rm m} \sim$ 1~GHz, presuming a driving frequency
$\omega$ in the order of 100~GHz.
The critical current is typically $I_{\rm c}\sim 0.1~\mu$A, which gives the order of bias current.

To summarize, we investigate theoretically a Landau-Zener-St\"{u}ckelberg interference of Majorana parity states in a junction between two one-dimensional topological superconductors
upon current injection. We demonstrate that the Majorana qubit can be rotated completely between the even- and odd-parity states, and a Landau-Zener-St\"{u}ckelberg interferometry
can be implemented by adjusting the pulse length of current injection, which along with the control of junction with gate voltage provides an arbitrary manipulation of the Majorana qubit.
The present proposal has the potential advantage that current injection can be performed quickly as compared with operation involving magnetic flux.
The scheme is scalable at least linearly upon introducing a sequence of one-dimensional topological superconductors coupled with quantum tunneling junctions. The dynamics of Majorana qubit upon current injection can be monitored by the power spectrum of microwave radiated from the junction.

\vspace{5mm}
\noindent\textbf{Acknowledgements}\\
This work was supported partially by the National Natural Science Foundation of China under Grants No. 11304400 and No. 61471401,
and partially by the WPI Initiative on Materials Nanoarchitectonics, Ministry of Education, Cultures, Sports, Science and Technology, Japan.


\begin{thebibliography}{00}

\bibitem{Majorana} E. Majorana, Nuovo Cimento \textbf{5} 171 (1937).
\bibitem{ReadGreen} N. Read and D. Green,  Phys. Rev. B \textbf{61}, 10267 (2000).
\bibitem{Kitaev} A. Y. Kitaev, Phys. Usp. \textbf{44}, 131 (2001).
\bibitem{ivanov} D. A. Ivanov, Phys. Rev. Lett. \textbf{86}, 268 (2001).


\bibitem{kanermp} M. Z. Hasan and C. L. Kane, Rev. Mod. Phys. {\bf 82}, 3045 (2010).
\bibitem{aliceareview} J. Alicea, Rep. Prog. Phys. 75, 076501 (2012).
\bibitem{beenakkerreview} C. W. J. Beenakker, Annu. Rev. Con. Mat. Phys. \textbf{4}, 113 (2013).
\bibitem{kanefu} L. Fu and C. L. Kane, Phys. Rev. Lett. {\bf 100}, 096407 (2008).
\bibitem{aliceanphy} J. Alicea, Y. Oreg, G. Refael, F. von Oppen and M. P. A. Fisher, Nature Physics \textbf{7}, 412 (2011).
\bibitem{Sau11} J. D. Sau, D. J. Clarke and S. Tewari, Phys. Rev. B \textbf{84}, 094505 (2011).
\bibitem{LiangWangHu12} Q.-F. Liang, Z. Wang and X. Hu, Eorophys. Lett. \textbf{99}, 50004 (2012).
\bibitem{WuLiangHu14} L.-H. Wu, Q.-F. Liang and X. Hu, Sci. Technol. Adv. Mater. \textbf{15}, 064402 (2014).


\bibitem{law09} K. T. Law, P. A. Lee and T. K. Ng, Phys. Rev. Lett. \textbf{103}, 237001 (2009).
\bibitem{sarma1d} R. M. Lutchyn, J. D. Sau and S. Das Sarma, Phys. Rev. Lett. {\bf 105}, 077001 (2010).
\bibitem{oppen1d} Y. Oreg, G. Refael and F. von Oppen,  Phys. Rev. Lett. \textbf{105}, 177002 (2010).
\bibitem{kouwenhoven} V. Mourik, K. Zuo, S. M. Frolov, S. R. Plissard, E. P. A. M. Bakkers and L. P. Kouwenhoven, Science {\bf 336}, 1003 (2012).
\bibitem{pergescience} S. Nadj-Perge, I. K. Drozdov, J. Li, H. Chen, S. Jeon, J. Seo, A. H. MacDonald, B. A. Bernevig and A. Yazdani, Science \textbf{346}, 602 (2014).




\bibitem{bonderson10} P. Bonderson, D. J. Clarke, C. Nayak and K. Shtengel, Phys. Rev. Lett. \textbf{104}, 180505 (2010).
\bibitem{hassler10} F. Hassler, A. R. Akhmerov, C. Y. Hou and C. W. J. Beenakker, New J. Phys. \textbf{12}, 125002 (2010).
\bibitem{jiang} L. Jiang, C. L. Kane and J. Preskill, Phys. Rev. Lett. {\bf 106}, 130504 (2011).
\bibitem{flengsberg11} K. Flensberg, Phys. Rev. Lett. \textbf{106}, 090503 (2011).
\bibitem{schmidt} T.L. Schmidt, A. Nunnenkamp and C. Bruder, Phys. Rev. Lett. 110, 107006 (2013).
\bibitem{pekker13} D. Pekker, C. Y. Hou, V. E. Manucharyan and E. Demler, Phys. Rev. Lett. \textbf{111}, 107007 (2013).





\bibitem{albrecht} S. M. Albrecht, A. P. Higginbotham, M. Madsen, F. Kuemmeth, T. S. Jespersen, J. Nyg\r{a}rd, P. Krogstrup and C. M. Marcus, Nature \textbf{531}, 206 (2016).

\bibitem{Fu} L. Fu, Phys. Rev. Lett. \textbf{104}, 056402 (2010).
\bibitem{WangHuLiangHu} Z. Wang, X.-Y. Hu, Q.-F. Liang and X. Hu, Phys. Rev. B \textbf{87}, 214513 (2013).

\bibitem{Josephson1} B.~D. Josephson, Phys. Lett.  \textbf{1}, 251 (1962).
\bibitem{Josephson2} B.~D. Josephson, Rev. Mod. Phys. \textbf{36}, 216 (1964).

\bibitem{landau} L. Landau, Phys. Z. Sowjetunion, \textbf{2}, 46 (1932)
\bibitem{zener} C. Zener, Proc. R. Soc. Landon Ser. A \textbf{137}, 696(1932)
\bibitem{stueckelberg} E. C. G. St\"ueckelberg, Helv. Phys. Acta \textbf{5}, 369 (1932)

\bibitem{leggett} A. J. Leggett, S. Chakravarty, A.T. Dorsey, M. P. A. Fisher, A. Garg and W. Zwerger, Rev. Mod. Phys. \textbf{59}, 1 (1987).

\bibitem{WangLZS} W. C.  Huang, Q. F. Liang, D. X. Yao and Z. Wang, Phys. Rev A \textbf{92}, 012308 (2015).

\bibitem{tinkham}  M. Tinkham, Introduction to Superconductivity (Second Edition, McGraw-Hill Book Co. 1996). 
\bibitem{husust} X. Hu and S.-Z. Lin, Supercond. Sci. Technol. \textbf{23}, 053001 (2010).

\bibitem{floquet} G. Floquet, Ann. Sci. Ec. Normale Super. \textbf{12}, 47 (1883).
\bibitem{shirley} J. H. Shirley, Phys. Rev. \textbf{138}, B979 (1965).
\bibitem{son} S.-K. Son, S. Han and S.-I. Chu, Phys. Rev. A \textbf{79}, 032301 (2009).


\bibitem{nori} S. N. Shevchenko, S. Ashhab and F. Nori, Phys. Rep. \textbf{492},1-30 (2010).
\bibitem{GGC} G. Cao, H. O. Li, T. Tu, L. Wang, C. Zhou, M. Xiao, G. C. Guo, H. W. Jiang and G.-P. Guo, Nature Communication \textbf{4}, 1401 (2013).


\end{thebibliography}
\end{document}